\documentclass[prb,preprint,showpacs,preprintnumbers,amsmath,amssymb]{revtex4}

\usepackage{graphicx}
\usepackage{dcolumn}
\usepackage{bm}

\begin{document}

\preprint{BMJ831}

\title{Paramagnetic Meissner effect depending on superconducting thickness 
in Co/Nb multilayers}

\author{S. F. Lee}
\author{J. J. Liang}%
\author{T. M. Chuang}%
\author{S. Y. Huang}%
\author{J. H. Kuo}%
\author{Y. D. Yao}%
\affiliation{%
Institute of Physics, Academia Sinica, Taipei, Taiwan 115, Republic of China
}%


\date{\today}

\begin{abstract}
Paramagnetic Meissner Effect (PME) was observed in Co/Nb/Co trilayers and 
multilayers. Measurements of the response to perpendicular external field 
near the superconducting transition temperature were carried out for various 
Nb thicknesses. PME was found only when layer thickness is no smaller than 
penetration depth of Nb. A classical flux compression model [Koshelev and 
Larkin, Phys. Rev. B \textbf{52}, 13559 (1995)] was used to explain our 
data. We inferred that the penetration depth was a critical length, below 
which superconducting current density became too small and the PME could not 
be achieved.
\end{abstract}

\pacs{74.60.Ec, 74.25.Ha, 74.60.Ge, 74.62.Bf}
\maketitle

%

\section{\label{sec:level1}Introduction}

Meissner effect is the intrinsic diamagnetic response of superconductivity 
to external magnetic field. Instead, Paramagnetic Meissner effect (PME), 
also known as Wohlleben effect \cite{1} refers to weak paramagnetic signals 
observed in some field-cooled superconductive samples. It was first observed 
in high T$_{C}$ ceramic samples and later on in Nb foils, a conventional 
superconductor (S). Although the experimental observations are similar, the PME in 
high T$_{C}$ cuprate and in BCS superconductors are attributed to different underlying physics (see below). What is common 
is that only certain samples show PME. Surface treatment of the Nb samples 
could make this effect disappear. \cite{2,3} Ferromagnetic (F) materials can 
suppress Cooper pairs very efficiently due to the exchange field. This 
proximity effect reduces T$_{C}$ near the S/F interface. Since the pioneer work of Hauser et al., \cite{15} it is 
well-know that ferromagnetic materials suppress T$_{C}$ of adjacent 
superconducting films much stronger than non-magnetic materials. 
Experimental data on Fe/Pb/Fe trilayers \cite{16} and on Nb/Fe 
multilayers \cite{17} showed that as S layer thickness decreased, T$_{C}$ 
dropped to zero at critical thickness 70 nm and 32 nm, respectively. In 
contrast, in Nb/Cu multilayers, \cite{18} superconductivity persisted down to 
Nb thickness less than 5 nm. Thus, Co modifies the 
interface properties much more pronounced than non-magnetic materials. We thus focus on the field-cooled magnetic moments in Co/Nb multilayers to 
study the interface effect and the role of penetration depth.

Paramagnetic field-cooled magnetization (FCM) for high T$_{C}$ cuprate 
samples was first reported by Svelindh \textit{et al.} \cite{4} using a non-commercial SQUID magnetometer, in which the samples were kept stationary during measurements. Zero-field-cooled magnetization of 
those samples showed usual Meissner effect, but FCM in small external fields 
were paramagnetic. The PME signal decreased with increasing external field 
and became usual diamagnetic one for large field. 

A systematic study of PME was later reported by Wohlleben and coworkers. \cite{1} They found several melt processed 
Bi$_{2}$Sr$_{2}$CaCu$_{2}$O$_{8}$ samples showed PME when field is $ \le 
$0.5 Oe. They invoked ``$\pi $-junction'' as a possible origin of the PME, 
which induced negative spontaneous supercurrents across weak links of 
crystalline structures. 

Results similar to the PME in High T$_{C}$ materials on Nb disks with perpendicular field were later reported by Thompson \textit{et al.} \cite{2} They showed PME disappeared by various surface treatments and suggested that 
the surface layers having strong flux pinning sites together with sample 
orientation and geometry were responsible for the observation of PME. The 
same conclusion was also reached by Kostic \textit{et al.} \cite{3} They showed that samples exhibit PME have surface T$_{C}$ different from their bulk values. Strong and reversible PME was reported to exist up to 2000 Oe by Pust \textit{et al.} \cite{6} Since the observed effect is reversible, flux motion was excluded from being responsible. However, Terentiev  \textit{et al.} \cite{20} reported that non-uniform applied fields or vortex dynamics play important roles in the PME in Nb films less than 100 nm thick.  Geim \textit{et al.} \cite{7} measured micrometer size Nb and Al disks by a ballistic Hall magnetometer with detection loops $\sim $1$\mu $m$^{2}$, which utilized two-dimensional electron gas in semiconductor 
heterostructure. They found PME to be an oscillating function of external 
field and concluded that the PME was related to surface superconductivity. 

Different theoretical models for the PME were proposed. One explanation was 
Josephson loops with a spontaneous negative critical current, i.e., $\pi 
$-junctions caused by impurities or grain boundary scattering. PME was also 
used as an evidence for d-wave symmetry in High T$_{C}$ materials. \cite{8} 
Since only in certain but not all samples can PME be found, it is most 
likely caused by surface or micro-structural defects rather than being an 
intrinsic property. To explain the PME in Nb, a flux compression model was 
demonstrated to result in 
paramagnetic moment because of inhomogeneous superconducting transition or 
the surface layer having a higher T$_{C}$ than the bulk material. \cite{9} It 
treated current and field distributions for the cases of complete and 
partial Bean state. Superconducting currents at the sample surface forming 
`giant vortexes' was raised as another model. \cite{10} From the 
self-consistent solution to the Ginzburg-Landau equations, it was proved 
that a giant vortex state at the surface can be formed when S is 
field-cooled at H$_{c3}$, the surface critical field. This state has an S 
order parameter with a fixed orbital quantum number $L$. When temperature 
decreases further, flux can be trapped and compressed into a giant vortex 
state with paramagnetic signal.

\section{Experimental methods}

We fabricated several series of Nb films, Nb/Co trilayers and multilayer 
samples by dc sputtering onto Si (100) substrates. Samples with 2$\mu $m 
line-width and 40$\mu $m in length were also fabricated by lithography 
technique to measure critical currents. Base pressure in the vacuum 
chamber is 2$\times$ 10$^{-7}$ Torr or better. Deposition was made 
under 1 mTorr Ar gas, with 0.05 nm/s rate for Co and 0.12 nm/s for Nb. 
Twelve samples can be fabricated in the same run. \cite{11,12} 
As the Nb target became thinner, the deposition rate increased slightly to 0.15nm/s. 
Prolong the pre-sputtering time between ignition of plasma and start 
of deposition of Nb improved the sample quality, as pointed out by 
Muhge et al. \cite{19} In this paper, we concentrate on three Nb 
thicknesses: 240nm, 80nm, and 30nm, with all samples having the same total 
Nb thickness in each series. Magnetic measurements at low temperatures were 
performed by a commercial SQUID magnetometer. Data presented here are 
measured with field perpendicular to layer plane. We have shown that the S 
penetration depth in our Nb/Co multilayers was about 30 to 40nm at each 
interface. \cite{12} The average saturation moment of Co decreased only when 
nominal thickness was less than 1.5nm. Nb layers with thickness of 30nm 
sandwiched between Co was just above the critical thickness, below which no 
S transition could be found. \cite{13} 

As already pointed out, \cite{5} S samples measured with commercial SQUID 
magnetometers showed artifact because the superconducting magnet has 
inevitably non-uniform field. Indeed, we reached the same conclusion that S 
signals depended on the sample position while cooling through T$_{C}$, on 
the scan length in the superconducting magnet, and on the field uniformity 
etc. Since the resulting signal, voltage versus position, was no longer 
symmetrical with respect to the magnet center, commercial fitting routine in 
the automated software sequence might sometimes pick the wrong maximum and 
give a wrong sign to the measured moment. We measured our sample with the 
samples kept stationary and with a temperature sweeping rate of 0.05K/min. 
Signals were normalized to a Pb foil sample measured below T$_{C}$ with a 
4-cm scan. Applied field was calibrated also by the Pb foil because of its 
large diamagnetic response. Moment versus field above T$_{C}$ was also 
measured by standard 4-cm scan with the SQUID voltage response monitored 
closely. Asymmetric curves were frequently observed, indicating non-ideal 
dipole response, due to the perpendicular direction was a hard axis. In cgs 
unit, the volume susceptibility for complete Meissner effect was --1/4$\pi 
$. The small total S volume of the samples, $\sim $3mm$\times $3mm$\times 
$240nm, would have made the signal to be detected below the SQUID limit. 
This was compensated by the demagnetization factor, d, of the perpendicular 
geometry. Taking into account this factor, magnetization per unit volume is 
given by:

\[
{\frac{{M}}{{V}}} = - {\frac{{1}}{{4\pi}} }{\frac{{1}}{{1 - d}}}H_{} .
\]

Calculation after Crabtree \cite{14} for thin cylinders of 3 mm diameter, 
240, 80, and 30nm height gave the geometry factors 1/(1-d) 
$\sim $3600, 7700, and 24000, respectively.

\section{Results and discussion}

Measurements on our pure Nb films with various thicknesses have not found 
any PME signal. Apparently the surface property of our films is not much 
different from the bulk. On the contrary, when the surfaces of the Nb films 
are covered by Co layers, PME was found when Nb thickness is larger than the 
penetration depth. 


Fig. 1 shows examples of field-cooled and zero-field-cooled magnetic 
response to perpendicular magnetic field versus temperature. Data were taken 
while warming. Clear paramagnetic signals were seen for these samples with 
Nb thickness 240 or 80 nm. The 80 nm sample showed larger signal than the 
240 nm one due to the demagnetization factor. When normalized to the ZFC 
susceptibilities, the signals correspond to 0.6 and 0.55 percents. 
However, several samples with 
30 nm thick Nb showed only Meissner effect. That is, PME was seen when Nb 
thickness was no less than the penetration depth. Since all samples have the 
same total Nb thickness, this behavior was not due to difference in S 
volume. Fig. 2 depicts a model for the occurrence of PME after Koshelev and 
Larkin. \cite{9} Since the free surface on the sides has higher T$_{C}$, 
superconductivity is established there first. Large amount of flux can be 
trapped inside the sample. The diamagnetic shielding current $I_{S}$ and 
paramagnetic pinning current $I_{P}$ are then functions of the sample 
geometry. For our thin plate orthogonal to the field, competition between 
the two currents results in the PME.


To be sure these PME signals were not due to the expelled magnetic flux from 
Meissner state of Nb layers enhancing the moments of the Co layers, we have 
studied the hysteresis loops at 10K with field perpendicular to the surface 
carefully. All our samples have easy axis for Co layers lying in the plane 
with coercive fields less than 80Oe. The perpendicular direction is a hard 
direction with much larger coercive field. The 
magnetic state at zero field depends on the history of the sample 
and can be manipulated by performing field minor loops to any point 
below remanence. The inset of Fig. 3 shows the hysteresis loop and two 
initial magnetization curves of the 240nm Nb sample. The slopes of the 
initial magnetization curve, i.e. the susceptibilities, indicate how easily 
the magnetic moments can be rotated out of the in-plane easy axis direction. 
These slopes for three samples are shown in Fig. 3. One can see that 
different Co thickness did not show much different dM/dH value at low field. 
The slope exhibited large change only when field was at 90 Oe or larger. 
Thus we rule out the possibility of the expelled flux lines when the Nb 
layers go into S state as the major contribution of the paramagnetic 
signals. 


Since the observed paramagnetic signals were indeed from the PME, we 
performed calculations for a complete Bean state following Ref. \onlinecite{9}. 
As an approximation, we use Eq. (13) in Ref. \onlinecite{9} for a 
disk sample with radius $R$ and thickness $d$. Because the presence 
of Co, moving magnetic flux around costs energy. Thus we assume a case 
of weak flux compression, i.e., the shielding current ($I_{S}$ in Fig. 2) flows 
only in the region $\lambda $ of several penetration depths near the free 
surface. The paramagnetic signal 
normalized to the complete Meissner signal is  \cite{9}

\[
{\frac{{M}}{{M_{M}}} } = 1.08{\left[ { - (1 - f) + {\frac{{\lambda 
}}{{R}}}\left( {\ln {\frac{{R}}{{\lambda}} } - 0.96} \right)} \right]}
\]

\noindent
where $f$ is the fraction of trapped flux. We obtained that the two samples in 
Fig. 1 both had $f$ around 99.5{\%}. If $\lambda $ is as long as 10 micron, $f$ 
is about 97{\%}. In a weak compression case, $f$ must be larger than 95{\%} 
and the PME is small. In a strong compression case, the PME signal ranges 
between 27{\%} and 0{\%} for $f$ ranges from 100{\%} to 80{\%}. For the dependence 
of resulting PME signals on $\lambda $ and $f$, see Fig. 4 in Ref. 13. 
The large $f$ in our case can be attributed to the presence of Co helps to 
trap flux and to reduce interface T$_{C}$.  The absence of PME in samples with 30nm 
thick Nb can be explained by the critical current density, $j_{C}$, and how 
fast $j_{C}$ increases with decreasing temperature. The critical current 
enters the above equation through $f$. The pinning current density required to 
maintain the compressed flux must not exceed $j_{C}$, otherwise the flux will 
move out of the sample and the PME cannot be achieved. Fig. 4 shows our data 
on $j_{C}$ versus temperature measured on 2 micron wide, 40 micron long 
samples. The 240 nm thick Nb sample has a value and shape close to the bulk 
behavior. The $j_{C}$ of the 30 nm sample is about 20{\%} of the 240 nm 
one close to T$_{C}$, that is, the slope of $j_{C}$ increasing with 
decreasing temperature differs by 5 times. Comparing the estimated pinning 
currents and the $j_{C}$'s, we found $j_{C}$ is more than two orders of 
magnitude larger in the 240 nm thick Nb sample. In the 30 nm Nb sample, the 
two currents are the same order. Thus, in the 30 nm sample, when temperature 
decreases from T$_{C}$, the increase of $j_{C}$ is insufficient to support 
the trapped flux. Paramagnetic flux has to move out of the sample and no PME 
can be formed. We infer that the penetration depth of S materials is a 
critical length due to the reduction of critical current density and of the 
slope of $j_{C}$ increasing with decreasing temperature. Only when S layer 
thickness is larger than the penetration depth, could the PME be found.


The role of proximity effect is essential in this study. It suppresses the 
interface T$_{C}$ to allow the free surface to go into superconducting state 
first while the sample is slowly cooling down. The critical current density 
near the interface is strongly reduced due to the fringe fields of strong 
magnetic Co layers.  A detail discussion of the proximity effect between 
Co and Nb will be published elsewhere. \cite{13} Here we discuss briefly 
the behavior of the related samples. From upper critical field measurements, we found the 
Ginzburg-Landau superconducting coherence lengths at zero degree were 12 nm 
to 13 nm parallel to the plane and 9 nm to 10 nm perpendicular to the plane. 
Close to T$_{C}$, the Ginzburg-Landau coherence length is inversely proportional to 
(T$_{C}$-T)$^{1 / 2}$. When Nb is 30 nm thick, the coherence length and 
penetration depth just below T$_{C}$ are as larger as the thickness. The 
fringe field of Co affects the thick Nb films only at the interfaces, but 
reduces the critical current for 30 nm Nb thickness a lot. Thus, the 
required amount of trapped flux in the model of Koshelev and Larkin cannot 
be fulfilled for 30 nm thick Nb samples.

\section{Summary}
Paramagnetic Meissner effect was studied for superconducting thin films with 
thickness in the regime close to penetration depth. We have shown clear 
evidence that Co/Nb/Co trilayers and multilayers showed PME when Nb 
thickness is no less than its penetration depth. The total magnetic moment 
is not linearly additive of each layer's due to the trapped flux and induced 
current are functions of sample geometry. This gives us the possibility to 
manipulate flux lines in engineered superconductor/ferromagnet structures.


\newpage 
\noindent
Figure Captions
 
Fig. 1. PME moment versus temperature of two samples: 
Co(15nm)/Nb(240nm)/Co(15nm) in circles and [Co(7.5nm)/Nb(80nm)]$\times 
$3/Co(7.5nm) in triangles. Samples were kept stationary and field cooled at 
H=0.57Oe. Measurements were made with a warming rate of 0.05K/Min. 
The normalized zero-field-cooled and field-cooled susceptibilities 
are shown in the inset with solid and dash lines. PME susceptibilities 
are 0.60 and 0.55 percents, respectively.
 
Fig. 2. Schematic cross-section of a Co/Nb/Co trilayer. The free surface has 
higher T$_{C}$ and superconducts first. Competition between 
diamagnetic shielding current, $I_{S}$, and paramagnetic pinning current, 
$I_{P}$, results in PME due to the thin plate orthogonal to field geometry. 
Drawing is not to scale.
 
Fig. 3. Slope of the initial magnetization curve of three samples: 
[Co(15nm)/Nb(240nm)] $\times $5/Co(15nm) in squares, 
[Co(7.5nm)/Nb(80nm)]$\times $15/Co(7.5nm) in circles, and 
[Co(2.2nm)/Nb(30nm)]$\times $40/Co(2.2nm) in triangles with perpendicular 
field measured at 10K. The first two samples do not have particular high 
value than the third one. This indicates that increased Co moment by 
repelled flux while Nb layers go into superconducting state is not 
responsible for the PME signals in Fig. 1. The inset shows the 
hysteresis loop of the first (Nb240nm) sample with two initial 
magnetization curves in square. The zero field states were reached by 
performing field minor loops. Notice the curves are close to 
linear at low field and the slopes are similar.
 
Fig. 4. Critical current density versus reduced temperature for Nb 
thicknesses 240 and 30 nm sandwiched between Co layers.

\end{document}